# Exploring the Academic Invisible Web

**Dirk Lewandowski**
Department of Information Science, Heinrich-Heine-University, Düsseldorf, Germany.
**Philipp Mayr**
Doctoral candidate at the Institute of Library and Information Science, Humboldt-University, Berlin, Germany

**Abstract**

**Purpose:** To provide a critical review of Bergman's 2001 study on the Deep Web. In addition, we bring a new concept into the discussion, the Academic Invisible Web (AIW). We define the Academic Invisible Web as consisting of all databases and collections relevant to academia but not searchable by the general-purpose internet search engines. Indexing this part of the Invisible Web is central to scientific search engines. We provide an overview of approaches followed thus far.

**Design/methodology/approach:** Discussion of measures and calculations, estimation based on informetric laws. Literature review on approaches for uncovering information from the Invisible Web.

**Findings:** Bergman's size estimate of the Invisible Web is highly questionable. We demonstrate some major errors in the conceptual design of the Bergman paper. A new (raw) size estimate is given.

**Research limitations/implications:** The precision of our estimate is limited due to a small sample size and lack of reliable data.

**Practical implications:** We can show that no single library alone will be able to index the Academic Invisible Web. We suggest collaboration to accomplish this task.

**Originality/value:** Provides library managers and those interested in developing academic search engines with data on the size and attributes of the Academic Invisible Web.

**Paper type:** Research paper

**Introduction**

Recent years demonstrate an unbroken trend towards end-user searching. Users expect search services to be complete, integrated and up-to-date. Educated users naturally want to retrieve the most comprehensive and largest index. But size is not the only issue. Even in the academic sector, where advanced search tools and dozens of relevant reference and full text databases are to be found, users to a large degree consult general-purpose Internet search engines to retrieve academic documents. Information professionals, who are used to tackling multiple data sources and varied, combined search environments, are forced to use oversimplified, general search engines.

The rise of Web search engines has brought with it some shifts in user behavior. Web search engines suggest that all information available can be searched within just one system. The search process itself is easy and highly self-explanatory. Within the last few years, professional information vendors (and libraries) have found that search engine technology can easily fit their needs for making academic content available for end-user searching. Keeping in mind that search engine technology is also widely used in a business context, it can be said that this technology is the new key concept in searching (see Lewandowski, 2006).

The reasons for this shift in information behavior are relatively clear. More and more scholarly content is provided exclusively on the web. The Open Access movement is only one current example for this paradigm change: from the traditional print publishing system to the electronic publishing paradigm. The consequence is a situation which Krause calls the poly-central information provision (Krause, 2003). A growing decentralization in the field of new information providers and changed user expectations and habits have led to a gap in the providing of information. General search engines take advantage of this gap. Google Scholar and Scirus show this very clearly: They do index parts of the Invisible Web, but unfortunately with results of questionable quality (see below). A recent review of existing technologies to index the Invisible Web can be found in Ru and Horowitz (2005). They identified the main problems and strategies in indexing the Invisible Web. According to Ru and Horowitz "indexing the web site interface" or "examining a portion of the contents" of an Invisible Web site are the two typical approaches.

The pivotal point in the dilemma is the Invisible Web (for a detailed discussion see Lewandowski, 2005b). Library collections and databases with millions of documents remain invisible to the eyes of users of general internet search engines. Furthermore, ongoing digitization projects are contributing to the continuous growth of the Invisible Web. Extant technical standards like Z39.50 or OAI-PMH (Open Archives Initiative – Protocol for Metadata Harvesting) are often not fully utilized, and consequently, valuable openly accessible collections, especially from libraries, remain invisible. It could be asked whether general-purpose search engines should pay more attention to the Invisible Web, but, as has been demonstrated in recent years, these seem to be lacking in terms of completeness and in-formation quality (see Mayr and Walter, 2005; Brophy and Bawden, 2005). So other institutions with experience in information organization should attend to this task.

The structure of this article is as follows: First, we discuss the competing definitions of the Invisible Web and give a definition for the Academic Invisible Web. Then, we retrace Bergman's study on the size of the Invisible Web, in which we find some serious errors. We suggest new approaches to determine a better size estimate. In the next part of the article, we discuss the approaches used so far to uncover information from the Invisible Web. In the discussion section, we offer implications as to how libraries should deal with the issue of the Academic Invisible Web and give a roadmap for further research on the topic.

**Defining the (Academic) Invisible Web**

In short, the Invisible Web is the part of the web that search engines do not add to their indices. There are several reasons for this, mainly limited storage space and the inability to index certain kinds of content. We discuss two definitions of the Invisible Web, where we do not distinguish between the Invisible Web and the Deep Web. Both terms are widely used for the same concept and using one or the other is just a matter of preference. We use the established term Invisible Web. Sherman and Price give the following definition for the Invisible Web:

> Text pages, files, or other often high-quality authoritative information available via the World Wide Web that general-purpose search engines cannot, due to technical limitations, or will not, due to deliberate choice, add to their indices of Web pages. (Sherman and Price, 2001, p. 57)

This is a relatively wide definition as it takes into account all file types and includes the *inability* of search engines to index certain content as well as their *choice* not to index certain types of content. In this definition, for example, spam pages are part of the Invisible Web because search engines choose not to add them to their indices.

Bergman defines this much more narrowly. Focusing on databases available via the web, he writes:

> Traditional search engines can not "see" or retrieve content in the deep Web – those pages do not exist until they are created dynamically as the result of a specific search. (Bergman, 2001)

Table 1 shows the different types of Invisible Web content according to Sherman and Price. It is easy to see that their view of the Invisible Web includes Bergman's view in the rows "content of relational databases" and "dynamically generated content."

Disconnected pages are a real problem of the Invisible Web, but to a lesser extent than with the surface web. If search engines could find these pages, there would be no problem indexing them. There is the technical problem of a lack of information about the existence of these pages.

Some other, more technical problems, such as dynamically generated pages and file types, have nearly been solved by now. It remains true that programs and compressed files are not readable for search engines, but this begs the question of what is the use of search engines being able to index these. Other file types mentioned by Sherman and Price, such as PDF, are read by all major search engines nowadays. But Flash and Shockwave content still remain a problem, due to the lack of sufficient text for the search engines to index. The main problem here lies in the inability of most search engines to follow links within flash sites.

Real-time content remains a problem because search engines cannot keep up with the rapid update rates of some sites. But in the current context of indexing the Academic Invisible Web, this content type can be left out. This also holds true for the other technical limitations described by Sherman and Price. Therefore, we think that efforts in indexing the Invisible Web in general, and the academic part of it in particular, should primarily focus on databases not visible to general search engines. Therefore, we stick to Bergman's definition of the Invisible Web. Particularly in the academic context, the content of databases is central. Technical limitations do not need to be taken into consideration for academic content, because it is mainly in formats such as PDF, which are technically readable by general-purpose search engines.

But not all limitations in indexing the Invisible Web are purely technical. Sherman and Price define four types of invisibility, where, for our purposes, the distinction between proprietary and free content is important. A large part of the Invisible Web relevant to academia is part of the Proprietary Web, mainly the content from publishers' databases.

From a library perspective, the Academic Invisible Web consists mainly of text documents (in different formats such as PDF, PPT, DOC). This is the content that libraries (or academic search engines) should add to their searchable databases to give the user a central access point to all relevant content.

Therefore, we define the Academic Invisible Web (AIW) as consisting of all databases and collections relevant to academia but not searchable by the general internet search engines.

In accordance with Lossau's claim that libraries need to discover the Academic Internet (Lossau, 2004), one could narrow the above definition to the content of the databases that should be indexed by libraries (using search engine technology). We do not intend to say that one library alone should make all content from the AIW visible in a search engine, but that libraries should follow a cooperative approach in making this content visible.

It should be kept in mind that the AIW is only one part of the Web relevant to libraries. The Academic *Surface* Web (ASW) contains a multitude of relevant documents as well, e.g. most Open Access Repositories are part of the surface web and can be crawled by general-purpose search engines without any problem. The study by Lawrence and Giles (1999) returned results showing that only about six percent of the indexable web are academic content.

| Type of Invisible Web Content | Why It's Invisible |
| --- | --- |
| Disconnected page | No links for crawlers to find the page |
| Page consisting primarily of images, audio, or video | Insufficient text for the search engine to "understand" what the page is about |
| Pages consisting primarily of PDF or Postscript, Flash, Shockwave, Executables (programs) or Compressed files (.zip, .tar, etc.) | Technically indexable, but usually ignored, primarily for business or policy reasons |
| Content in relational databases | Crawlers cannot fill out required fields in interactive forms |
| Real-time content | Ephemeral data; huge quantities; rapidly changing information |
| Dynamically generated content | Customized content is irrelevant for most searchers; fear of "spider traps" |

Table 1: Types of Invisible Web Content (Sherman and Price, 2001, p. 61)

The AIW is valuable for scholars, librarians, information professionals and all other academic searchers and can provide everything relevant to the scientific process. This includes:
- literature (e.g. articles, dissertations, reports, books)
- data (e.g. survey data)
- pure online content (e.g. Open Access documents)

The main institutional providers of AIW content are:
- Database vendors, producing bibliographic metadata records enriched by human subject indexing (thesauri, classifications and other knowledge organization systems) and additional services like document delivery
- Libraries, also producing bibliographic records in openly accessible systems like Online Public Access Catalogues (OPACs), offering their collections enriched by human subject indexing and additional services
- Commercial Publishers, providing mainly full text content
- Other repositories of societies and corporations (e.g. the Association for Computing Machinery)
- Open Access repositories (e.g. Citebase, OpenROAR)

A lot of these materials are not necessarily part of the AIW, but are in fact uncovered by the main search engines and tools. For users of these heterogeneous collections, this means becoming accustomed to the respective systems and infor-mation structures. For example, most providers of scholarly information maintain their own subject access and information organization models, due to various traditions and indexed content types. Libraries index mainly books and compilations with their standardized universal authority files; database producers use proprietary domain-specific thesauri and classifications for indexing journal articles, while publishers use a mixture of manual and automatic indexing for their full texts. This results in a heterogeneity (Krause, 2003) between the collections and a complex situation for users in need of cross-database searching.

**Measuring the size of the (Academic) Invisible Web**

To our knowledge, the only attempt to measure the size of the Invisible Web was Bergman's study (2001). The main findings were that the Invisible Web is about 550 times larger than the surface web and consists of approximately 550 billion documents. Bergman's paper is widely cited and therefore we will discuss it in detail. Most other studies use Bergman's size estimates or estimate the size of the Invisible Web based on the ratio between surface and Invisible Web of 1:550 given by Bergman (e.g. Lyman *et al.*, 2003).

The basis for Bergman's size estimates is a "Top 60" list containing the largest Deep Web sites. These are put together manually from directories of such sites, while duplicates are removed. Bergman's Top 60 contains 85 billion documents with a total size of 748,504 GB. The top two alone contain 585,400 GB, which is more than 75 percent of the Top 60 (file size measure).

A further assumption is that there are around 100,000 Deep Web databases. This number comes from an overlap analysis between the largest directories of Invisible Web sites. Bergman's further calculations use the mean size of 5.43 million documents per Invisible Web database. Therefore, he states that the total size of the Invisible Web (mean multiplied by the number of databases) is 543 billion documents. Bearing in mind that the size of the surface Web at the time of the investigation (2001), was approximately 1 billion documents (based on data from Lawrence and Giles, 1999), Bergman finds that the Invisible Web is 550 times larger than the surface web.

These numbers were soon challenged (Sherman, 2001; Stock, 2003), but these authors just made new guesses and did not deliver a new calculation or even an explanation as to why Bergman's figures had to be mistaken. Our investigation found that the error lies in the *use of the mean* for the calculation of the total size estimate. While the mean is very high, the median of all databases is relatively low with just 4,950 documents. Looking at Bergman's Top 60 list, we see that the distribution of database sizes is highly skewed (Fig. 1), so the mean cannot be used to calculate the total size.

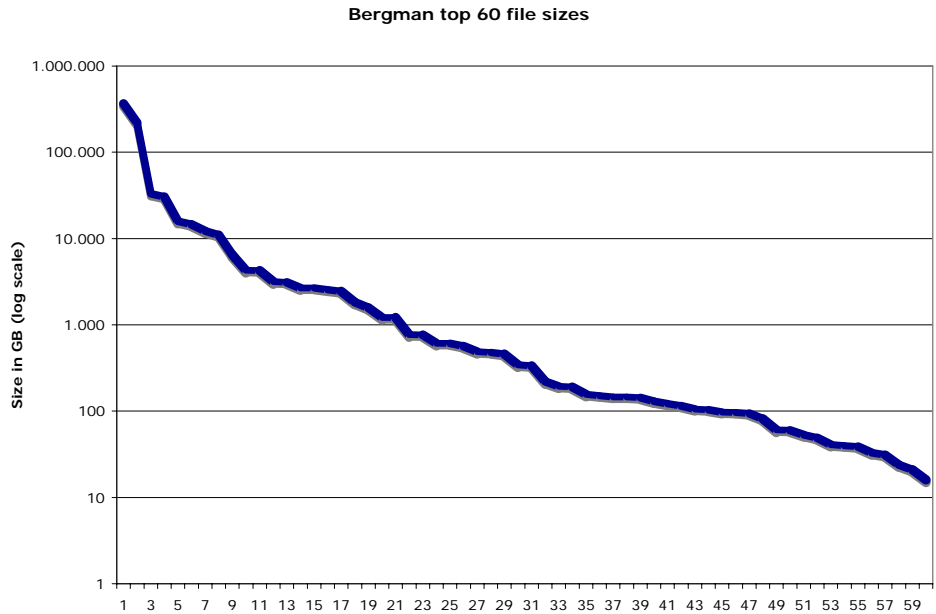

Figure 1: Distribution of file sizes in Bergman's Top 60

The skewed distribution of database sizes is typical and can also be seen in other database portfolios such as the DIALOG databases accessible via the Web. Again we see a highly skewed distribution (long tail). The sizes of the 347 files in DIALOG are plotted along a logarithmic scale (see Figure 2), demonstrating that there are few databases with more than 100,000,000 records (compare to Williams, 2005), and the majority with less than 1,000,000 records. The distribution is described by an exponential function with a high Pearson correlation (Pearson is 0.96, see Figure 2). The median of all 347 database sizes is circa 380,000 records. We hypothesize that the AIW will also follow such an exponential distribution.

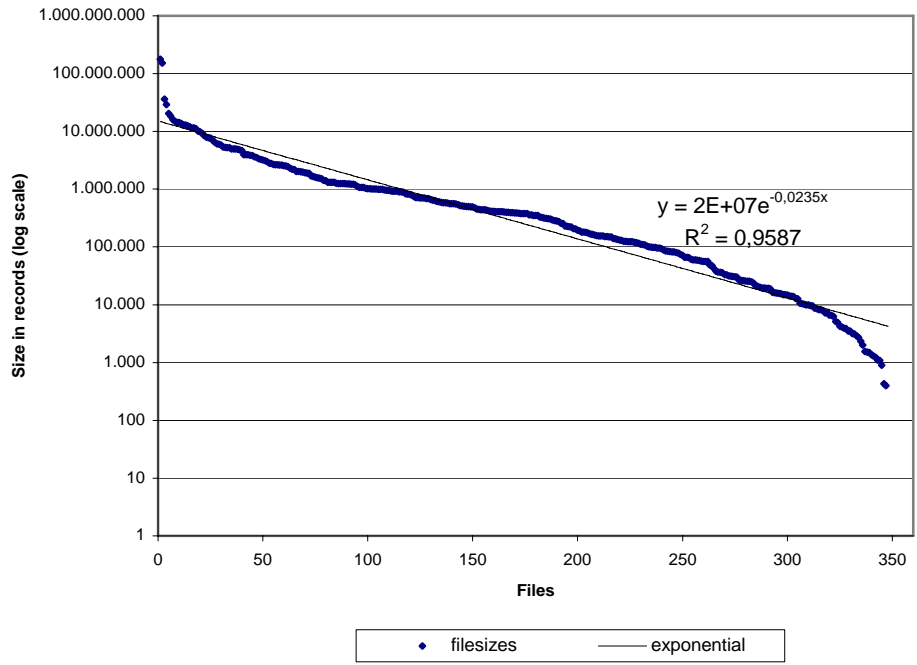

Figure 2: Distribution of database sizes from the DIALOG host (n=347)

For all further calculations in Bergman's study, the *size in GB* is used instead of the number of records per database. This is very problematic, as it is impossible to derive the record counts from the file size data due to the greatly varying size of database records (pictures, bibliographic records, full text records). Therefore, we are not able to make a more accurate calculation from Bergman's data. We can say that his size estimates are far too high, because of two fundamental errors. Firstly, the statistical error of using the mean instead of the median calculation, and secondly his misleading projection from the database size in GB. When using the 85 billion documents from his Top 60, we can assume that the total number of documents will not exceed 100 billion because of the highly skewed distribution. Even though this estimate is based on data from 2001, we think that the typical growth rate of database sizes (cf. Williams, 2005) will not affect the total size to a large extent.

But how much of the Invisible Web is academic content? Looking at Bergman's Top 60, we find that indeed 90 percent can be regarded as academic content, but if one chooses to omit all databases containing mere raw data, the portion of academic content shrinks to approximately four percent (Fig. 3), which corresponds to the amount of academic content found on the Surface Web in the 1999 study from Lawrence and Giles. The main part of Bergman's Invisible Web consists of raw data, mainly pictures such as satellite images of the earth. The records of these databases are far bigger than those of textual databases. Because Bergman only uses GB sizes, one cannot calculate new size estimates based on record numbers from the given data. For this task, one needs to build a new collection of the biggest Invisible Web databases.

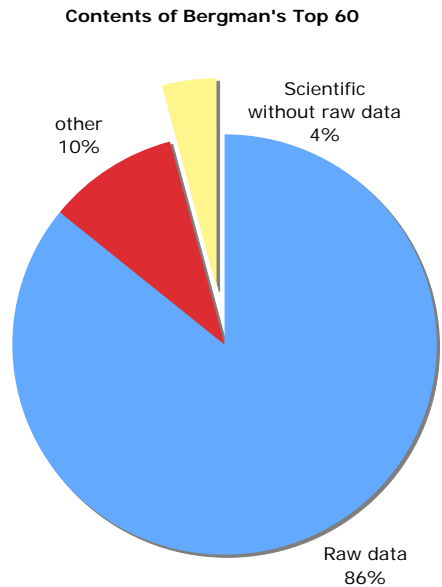

Figure 3: Contents of Bergman's Top 60

In summary, Bergman's study exhibits shortcomings in the mixture of database types and database content, as well as the calculation method used. It goes beyond the scope of this paper to present an exact size estimate for the Invisible Web. Further research is needed. In particular, a reliable collection of the largest Invisible Web databases should be built.

As we are not satisfied with Bergman's size estimates or our own raw estimate from Bergman's data, we have used additional data from the Gale Directory of Databases (Williams, 2005) for comparison. The directory contains approximately 13,000 databases and covers all major academic databases, as well as a number of databases solely of commercial interest. The total size estimate for all databases is 18.92 billion documents. The average size per database is 1.15 million records, with a highly skewed distribution. Five percent of the databases contain more than one million records, some more than 100 million. Omitting these very large databases, the mean database size is about 150,000 records. The total size estimate is calculated by adding the known database sizes and assuming the mean of 150,000 records for every other database. This method only works when all the very large database sizes are known. We cannot verify if all these are considered, but we found that some of the databases included in Bergman's Top 60 are missing from the Gale Directory. Therefore, it is difficult to compare the numbers directly. Because of the missing databases, the numbers from Gale are probably too low. In conclusion, we can only make an educated guess as to the actual size of the AIW: In our opinion, its size lies between 20 and 100 billion documents, viewing the raw data as part of the AIW. If these data were to be omitted, the AIW would be far smaller. What we can definitely say, is that the size of the AIW lies within the range of the index sizes of the biggest surface web search engines (Lewandowski, 2005c). Therefore, the challenge in indexing the whole AIW can only be met through a cooperative effort and not by a single institution acting alone.

Williams (2005) divides the databases in the Gale directory into six classes: word-oriented, number-oriented, image/video, audio, electronic services and software. For libraries and academic search engines, it is mainly word-oriented databases, comprising about 69 percent of all databases, which are of interest. Of these 8,994 word-oriented databases, some 80 percent are full-text or bibliographic information. We feel that these numbers represent a good starting point when attempting to index the whole Academic Invisible Web.

**Approaches to indexing and opening the Academic Invisible Web**

There are different models for enhancing access to the AIW, of which we can mention only a few. The four systems to be described shortly have a common focus on scholarly information, but the approaches and the content they provide are largely different. Google Scholar and Scirus are projects started by commercial companies. The core of their content is based on publishers' repositories plus openly accessible materials. On the other hand, Bielefeld Academic Search Engine (BASE) and Vascoda are academic projects where libraries and information providers open their collections, mainly academic reference databases, library catalogues plus free extra documents (e.g. surface web content). All systems use or will use search engine technology enhanced with their own implementations (e.g. citation indexing, specific filtering or semantic heterogeneity treatment).

*Google Scholar* [1] is currently the most discussed approach (Notess, 2005). The beta version, online since November 2004, covers some million documents. Google Scholar indexes a substantial part of international STM (Science – Technology – Medicine) publishers and other publishers who joined from the Crossref initiative. Google set up a prototype with great potential, but which also exhibits some unwelcome characteristics (Lewandowski, 2005a; Mayr and Walter, 2005). To its credit, Google Scholar tries to adopt the influential citation measure introduced by the Institute of Scientific Information (ISI) and implemented in the former Science Citation Index, now Web of Science. Unfortunately, Google Scholar provides no documentation that would make the service more transparent (Jacsó, 2005). It is impossible to say anything about the exact coverage, or how up-to-date the current service is, as a recent empirical study shows (Mayr and Walter, 2005).

*Scirus* [2] (see "Scirus White Paper", 2004) is a scientific search engine that indexes the academic surface web and also several other collections such as Elsevier's Science Direct and open access sources. This approach comes close to the desired combination of surface web content and AIW content, but is far from being complete, at least in the AIW part. With approximately 250 million documents from the surface web, Scirus is by far the largest search engine of its kind built with FAST technology (McKiernan, 2005).

*BASE* [3] (see Lossau, 2004) is an integrated search engine combining data from the library catalogue of Bielefeld University Library and data from approximately 160 open access sources (more than 2 million documents). It uses the FAST search engine.

*Vascoda* [4] is the prototype of an interdisciplinary science portal integrating library collections, literature databases and additional scholarly content. Vascoda acts as a meta portal delegating requests to lower, domain-specific layers or clusters. Each domain is responsible for its own subject portal which can be built using various technologies. Vascoda is an alternative model for a system bridging the gap of the AIW, designed by German libraries and documentation centers. Vascoda will soon launch its latest version enhanced by FAST search engine technology.

The roundup of these prototypical academic search systems shows clearly that serious efforts to index the AIW will need a collaborative approach. Every single approach has its own specific strengths and weaknesses. On the one hand we see broad cover-age with a bias towards commercial hits and the inability to exclude non-academic records from the results. On the other hand, we have limited scope and a lack of full text information.

**Discussion and conclusion**

Search engines are increasingly acquiring a gatekeeper function and are widely seen as offering general access to information due to their simplicity, search velocity and broad coverage. But this is true only for a part of the web.

As called for by Lossau (2004), libraries should discover the Academic Web. Although we focused on the Academic Invisible Web, there are also parts of the visible Web relevant to libraries. The key in achieving the best experience for the library user lies in a combined approach for both types of content. We were able to show that the AIW is very large and that its size is comparable to the indices of the largest general-purpose Web search engines. Therefore, only a cooperative approach is possible.

We conclude that existing search tools and approaches show potential to make the AIW visible. What we do not see is a real will for lasting collaboration among the players mentioned. Commercial search engine providers with their technological and financial superiority should work together with libraries, which have long experience in collection building and subject access models. They developed complex instruments for information organization (e.g. thesauri, classifica-tion, taxonomies) which could be highly valuable for end-user searching, automatic indexing, ontology building and classification of academic content. Publishers and database vendors should join by opening their collections (see Google Scholar example).

Unfortunately, we were not able to give a more precise size estimate for the Academic Invisible Web. Further research should focus on this task. We need to build a collection of the largest AIW databases and use the informetric distribution which we assume to be also given for the AIW. A good size estimate could be given based on such a sample.

Another task is to classify the AIW content to get a picture of the extent to which the different disciplines contribute to its size. Recommendations as to how to build specialized search engines for the various disciplines could be given based on such a classification.

A final research task is the distinction between the Visible and the Invisible Web. In the past years, we saw the conversion of large databases into HTML pages for the purpose of becoming indexed by the main Web search engines. Although this is mainly done in the commercial context, some libraries followed this approach with varying degrees of success (cf. Lewandowski, 2006). If database vendors make their databases available on the visible Web, libraries could follow the approach of Google or other search engines in indexing this content. Further research on this topic is needed, because at the current time nobody knows to what extent database content is already available on the surface web.

We can further conclude that Bergman did a good job in bringing the topic of the Invisible Web into the discussion, but, as we can demonstrate, his calculation is misleading for academic text-based content.

**Notes**

[1] http://scholar.google.com/
[2] http://www.scirus.com
[3] http://www.base-search.net/
[4] http://www.vascoda.de/